\documentclass[11pt]{amsart}
\usepackage{amsfonts}
\usepackage{amsmath,amsthm}
\usepackage{latexsym}
\usepackage{array}
\usepackage{amssymb}
\usepackage{enumerate}
\usepackage{comment}
\usepackage{tikz}

\DeclareFontFamily{U}{wncy}{}
\DeclareFontShape{U}{wncy}{m}{n}{<->wncyr10}{}
\DeclareSymbolFont{mcy}{U}{wncy}{m}{n}
\DeclareMathSymbol{\Sh}{\mathord}{mcy}{"58}

\usepackage[francais,english]{babel}
\usepackage{color}
\usepackage[latin1]{inputenc}

\theoremstyle{plain}
\newtheorem{theorem}{Theorem}[section]
\newtheorem*{theorem*}{Theorem}

\newtheorem{problem}[theorem]{Problem}

\theoremstyle{remark}

\newtheorem*{lem*}{Lemma}
\newtheorem*{sublem*}{Sublemma}
\newtheorem*{remark*}{Remark}
\newtheorem*{NB*}{NB}

\newcommand{\R}{ \mathbb{R} }

\newcommand{\Z}{ \mathbb{Z} }

\newcommand{\T}{ \mathbb{T} }

\newcommand{\gF}{\mathfrak F}

\newcommand{\ssum}{L^{-d}\sum}

\newcommand{\cY}{ \mathcal{Y} }

\newcommand{\cF}{ \mathcal{F} }

\newcommand{\EE}{ {\mathbb E}}

\newcommand{\cT}{ \mathcal{T} }

\newcommand{\I}{ \mathcal{I} }

\newcommand{\om}{ \omega }

\newcommand{\ga}{\gamma }

\renewcommand{\phi}{ \varphi }
\newcommand{\oms}{ \omega^{12}_{3s}  }

\newcommand{\eps}{\varepsilon}

\newcommand{\de}{ \delta }
\newcommand{\al}{ \alpha }

\newcommand{\dess}{\delta'^{12}_{3s}}
\newcommand{\des}{\delta^{12}_{3s}}

\newcommand{\ben}{\begin{equation*}}
\newcommand{\een}{\end{equation*}}

\newcommand{\be}{\begin{equation}}
\newcommand{\ee}{\end{equation}}

\numberwithin{equation}{section}

\newcommand{\ov}{ \overline }

\newcommand{\lan}{ \langle }

\newcommand{\ran}{ \rangle}

\newcommand{\lbl}{\label}

\newcommand{\qu}{\quad}
\newcommand{\qmb}{\quad\mbox}

\newcommand{\volna}{\thicksim}

\newcommand{\Ga}{\Gamma}

\title{On the Zakharov-L'vov stochastic model for
	wave turbulence}
	
\author{Andrey  Dymov}
\address{Steklov Mathematical Institute of RAS, Moscow 119991, Russia 
	\& National Research University Higher School of
	Economics, Moscow 119048, Russia} \email{dymov@mi-ras.ru}
\author{Sergei Kuksin}
\address{Universit\'e Paris-Diderot (Paris 7), UFR de Math\'ematiques - Batiment Sophie Germain, 5 rue Thomas Mann, 75205 Paris,
 France  \& School of Mathematics, Shandong University, Jinan, PRC \& Saint Petersburg State University, Universitetskaya nab., St. Petersburg, Russia}
\email{ Sergei.Kuksin@imj-prg.fr}

\date{}
\begin{document}
\maketitle

\begin{abstract}
		In this note we present the main results of the papers \cite{DK, DK2}, dedicated to   rigorous study of the limiting
properties of the stochastic model for  wave turbulence due to  Zakharov-L'vov. Proofs of the assertions, stated below
		without reference, may be found in those works. 
	\end{abstract}
	
	\medskip
	

	\section{Introduction}\label{s2}
	\subsection{The model}
	Let $L\geq 1$ and ${\mathbb{T}}^d_L={  \mathbb{R}   }^d/(L{  \mathbb{Z}   }^d)$ be a torus of dimension $d\geq 2$. 
	Denote by  $H$ the space  $ L_2({{\mathbb T}}^d_L;{{\mathbb C}})$   with respect to the
	normalised Lebesgue measure:
	$$
	\|u\|^2 =\|u\|^2_{L_2(\T^d_L)}=
	\lan u, u\ran\,,
	\quad
	\lan u, v\ran =
	L^{-d} \Re\int_{{{\mathbb T}}^d_L} u \bar v\,dx\,.
	$$
	Consider the modified NLS equation
	$$
	\frac{{{\partial}}}{{{\partial}} t}  u +i\Delta u-   i\nu \, \big( |u|^2 - \|u\|^2\big) u = 0\,,
	\qquad \Delta = {(2\pi )^{-2}} 
	\sum_{j=1}^d ({{\partial}}^2 / {{\partial}} x_j^2)\,,\;\;x\in\T^d_L,
	$$
	where $\nu\in(0,1/2]$. This is a hamiltonian PDE, obtained by modifying the standard cubic NLS
	equation by another hamiltonian equation 
	$
	\frac{{{\partial}}}{{{\partial}} t}  u  = -i\nu  \|u\|^2 u,
	$
	whose flow commutes with that of the cubic NLS. The modified NLS can be obtained from the standard cubic equation  by the substitution
	$
	u = \exp{( i t\nu\| u'\|^2)} u'.
	$
	 This modification is rather often used by people, working on hamiltonian PDEs since it
	keeps the main features of NLS, reducing some non-critical 	  technicalities. 
	The role of the modification is to remove from the Hamiltonian of the NLS its integrable part (see below the second footnote). 
	 If instead the cubic equation  we considered the quadratic NLS,
	corresponding to a three--wave  system,  the modification would not be needed.

	 We will write solutions $u$ as $u(t,x)\in {{\mathbb C}} $ or as $u(t)\in H$. 
	Passing to the slow time $\tau = \nu t$ we re-write the equation as
	\begin{equation}\label{ku1}
	\dot u +i \nu^{-1} \Delta u -   i \, \big( |u|^2 - \|u\|^2\big) u=0\,,\quad \dot u = ({{\partial}}/{{\partial}} \tau)u(\tau, x),\;\; x\in\T^d_L \,.
	\end{equation}
	The objective  of  Wave Turbulence (WT) is to study solutions of \eqref{ku1} when
	\be\label{limit}
\nu\to0, \quad L\to\infty.
\ee

	We will write the Fourier series for $u(x)$ in the form
	$$
	u(x)= L^{-d/2} {\sum}_{s\in{{\mathbb Z}}^d_L} v_s e^{2\pi  i s\cdot x},\qquad {{\mathbb Z}}^d_L = L^{-1} {{\mathbb Z}}^d\,,
	$$
	where
	$\ 
	v_s := \hat u(s) = L^{-d/2}\int_{\T^d_L} u(x) e^{-2\pi i s\cdot x}\,dx \,.
	$
	Then
	\be\label{norm}
	\|u\|^2 = {\ssum}_s |v_s|^2 =: \|v\|^2= \|v\|^2_{L_2(\Z^d_L)}\,.
	\ee
	 Denote by $h$ the Hilbert space 
	$h=\big(L_2(\Z^d_L), \|\cdot\|\big)$. The Fourier transform defines an isomorphism
	$\ 
	H\to h,\quad u(x) \mapsto (v_s=\hat u(s))\,.
	$
		
	 When  studying  eq. \eqref{ku1}, people from the WT community often talk
	about ``pumping   energy to low modes  and dissipating it in  high modes".
	To make this rigorous,  Zakharov-L'vov \cite{ZL75} (also see in \cite{CFG}, Section~1.2) 
	suggested to consider the NLS equation, dumped by a (hyper)viscosity and driven by a random force:
	\begin{equation}\label{ku3}
	\dot u +   i \nu^{-1} \Delta u -   i\rho\,( |u|^2 -\|u\|^2) u 
	 =- (-\Delta +1)^{r_*} u + \dot \eta^\omega(\tau,x), \quad r_*>0,
	\end{equation}
	where $\rho>0$, and the random process $\eta$ is given by its Fourier series
	\be\lbl{eta}
	\eta^\omega(\tau,x)=
	L^{-d/2} {\sum}_s b_s \beta^\omega_s(\tau) e^{2\pi  i s\cdot x}.
	\ee
	Here $\{\beta_s(\tau), s\in{{\mathbb Z}}^d_L\}$ are standard independent complex Wiener processes~\footnote{
		i.e. $\beta_s = \beta_s^1 + i\beta_s^2$, where $\{\beta_s^j, s\in {{\mathbb Z}}^d_L, j=1,2  \}$ are
		standard independent real Wiener processes. 	}, 
the constants $b_s>0$ fast decay when $|s|\to\infty$ and are obtained as the restriction to ${{\mathbb Z}}^d_L$ of a positive Schwartz  function on ${{\mathbb R}}^d\supset \Z^d_L$.
 It is known that if $r_*$ is sufficiently large then the Cauchy problem for equation \eqref{ku3} is well-posed. 
Applying the Ito formula to   \eqref{ku3} and denoting  $B =  {\ssum}_s b_s^2$
	we get the  balance of energy for solutions of eq.~\eqref{ku3}:
	\be\label{balance}
	{{\mathbb E}} \|u(\tau)\|^2 +2{ {\mathbb E} } \int_0^\tau  \| (-\Delta+1)^{r_*} u(s)\|^2 \, ds ={{\mathbb E}} \|u(0)\|^2 +2B \tau\,.
	\ee
	We see that  ${{\mathbb E}} \|u(\tau)\|^2$ -- the averaged energy per volume of a
	 solution $u$ -- is of  order one uniformly in $L$, no matter how big or small $\rho$ is. 
Later $\rho$ will be scaled with $\nu$ in such a way that an equation for the distribution of solution's energy along  the spectrum which
follows from eq.~\eqref{ku3} admits a non-trivial kinetic limit. As we will see this  requirement determines the scaling of $\rho$ 
 uniquely.
	\medskip

	Passing to the Fourier presentation, we write \eqref{ku3} as
	$$
	\dot v_s -i\nu^{-1}|s|^2v_s + \gamma_s v_s
	=  i\rho  L^{-d}  {\sum}_{1,2}  \dess v_{1} v_{2} \bar v_{3}+b_s \dot\beta_s \,, \quad s\in{{\mathbb Z}}^d_L,
	$$
	where   $\gamma_s = (1+|s|^2)^{r_*}$. Following the tradition of WT we 
	abbreviate $v_{s_j}$ to $v_j$, $\ga_{s_j}$ to $\ga _j$ etc, abbreviate 
	$\sum_{s_1, s_2\in \Z^d_L} $ to $\sum_{1, 2} $ 
	 and denote \footnote{If eq. \eqref{ku1} is replaced by the standard NLS equation, then 
	$\dess$ should be modified to $\des$, obtained by dropping in the definition of $\dess$
	 the condition $ \{s_1, s_2\} \ne \{s_3, s\}$. Then the double sum 
	in the $v$--equation  will be modified by adding the ``integrable term" 
	$i\rho L^{-d} (2v_s\sum_m |v_m|^2 -v_s|v_s|^2)$. 
	}
$$
  \dess =
   \left\{\begin{array}{ll} 1,&
 \text{if $s_1+s_2=s_3+s$ and }  \{s_1, s_2\} \ne \{s_3, s\}
 \\
 0,& \text{otherwise}.
 \end{array}\right.
$$
 In  view of the factor  $\dess$,
	in  the double  sum above  $s_3$ is a function of $s_1, s_2, s$, i.e. $s_3 =s_1+s_2-s$. 
	Using the interaction representation
	$$
	v_s(\tau) = \exp({  i \nu^{-1}\tau |s|^2} ) \, a_s(\tau)\,, \;\; s\in {{\mathbb Z}}^d_L\,,
	$$
	we re-write the $v$--equations   as  $a$--equations:
	\begin{equation}\label{ku4}
	\begin{split}
	&\dot a_s + \gamma_s a_s
	=  i\rho \cY_s(a;\nu^{-1}\tau) +b_s \dot\beta_s  \,,\quad s\in{{\mathbb Z}}^d_L\,,\\
	&\cY_s(a; t )=L^{-d}  {\sum}_{1,2}  \dess  a_{1} a_{2} \bar a_{3}
	e^{  i t \omega^{12}_{3s}}.
	\end{split}
	\end{equation}
	Here		$\{\beta_s\}$ is another set of  standard independent complex Wiener processes 
	and
	$$
	\omega^{12}_{3s}= |s_1|^2+|s_2|^2-|s_3|^2 - |s|^2 = -2(s_1-s)\cdot (s_2-s)\,
	$$
	(the last equality holds since  $s_3=s_1+s_2-s$ in view of the  factor $\dess$).
	By $\cY_s(a^1, a^2, a^3; t )$ we will denote the natural
	poly-linear mapping, corresponding to the 
	3-homogeneous mapping $\cY_s$; so $\cY_s(a; t ) = \cY_s(a,a,a; t )$.

	\subsection{Background}

	The {\it energy spectrum} of a solution $u(\tau)$ of eq.~\eqref{ku3} is the function
	$$
	{{\mathbb Z}}^d_L \ni s\mapsto n_s(\tau) = n_s(\tau; \nu,L)
	=   {{\mathbb E}} |v_s (\tau) |^2 =  {{\mathbb E}}|a_s (\tau)|^2.
	$$
Traditionally in the center of attention for people, working on WT, 
 is the limiting behaviour of the function $n_s(\tau) $ and of  correlations of solutions $a_s(\tau)$
under the limit \eqref{limit}. 
One of the main predictions of WT is that under this limit the energy spectrum $n_s(\tau)$ satisfies a
{\it wave
kinetic equation} (WKE).
There are plenty of physical works,  containing different (but consistent) approaches to the study of the energy spectrum $n_s$ under the limit \eqref{limit} 
and to the derivation for it   the WKE 
(e.g. see \cite{ZLF,Naz11, NR} and references therein; see also introduction to the work \cite{BGHS}). 
Non of them was ever rigorously justified despite the strong interest in  mathematical community to the questions, addressed by these works.

Exact meaning of the limit \eqref{limit}  is not quite clear. It is known (see in \cite{KM}) that for $\rho$ and $L$
fixed, eq.~\eqref{ku4} has a limit as $\nu\to0$, and it was demonstrated in \cite{KM} on the physical level of
rigour that if we scale $\rho$ as $\tilde\eps\sqrt L$, $\tilde\eps\ll1$, then the iterated limit $L\to\infty$ leads to the WKE. Attempts to justify the latter result   rigorously so far have failed.

There are only a few rigorous  works, addressing the limit \eqref{limit}.
In the paper \cite{FGH} the authors  consider the deterministic NLS equation, take $d=2$ and calculate the 
limit \eqref{limit}  in a regime, when $L$ goes to infinity 
much slower than $\nu^{-1}$. The obtained 
 elegant description of the limit
 is far from the prediction of  WT and rather should be regarded as a kind of averaging. 
In the recent  paper \cite{BGHS} the authors study the deterministic NLS equation with random initial data $u(0,x)$ and 
choose   the phases $\arg v_s(0)$, $s\in\Z^d_L$, of the Fourier coefficients of $u(0,x)$ to be independent  uniformly distributed  random variables. In the notation of our work they 
prove that under the limit \eqref{limit}, if $L$ goes to infinity slower than $\nu^{-1}$ but not too slow, then for the values of the slow time 
 $\tau$ of order $L^{-\de}$, $\de>0$, the energy spectrum $n_s(\tau)$ approximately satisfies a linearisation in time at $\tau=0$ 
   of the WKE, with the wave kinetic integral multiplied by $\nu$.  

Another problem of this kind was rigorously treated in \cite{LS}, where to achieve a progress in the study  of the deterministic NLS equation
 the authors had to replace the space--domain
$\T_L^d$ by the discrete torus $\Z^d/ (L\Z^d)$ and to modify the discrete Laplacian on  $\Z^d/ (L\Z^d)$ to a suitable operator, diagonal in the Fourier 
basis. The randomization was introduced through the initial data by assuming that $u(0,x)$ is distributed accordingly to the Gibbs measure of the system, 
 so the solution $u(t)$ is a stationary in time random process in $H$. A related problem is treated in  \cite{Faou}. 

\subsection{Results}
 In
this work we specify the limit \eqref{limit}  as follows:
\be\label{assumption}
\begin{split}
\nu\to0 \text{ and } L\ge\nu^{-2-\epsilon} \text{ for some }  \epsilon>0,\\
\text{or first $L\to\infty$ and next $\nu\to0$}.
\end{split}
\ee
The second option formally corresponds to the first one with $\epsilon=\infty$. 
This well agrees with a postulate widely accepted in the physical community that to get a kinetic limit, $L$ should go to infinity very fast while $\nu^{-1}$~--- not so fast.

We supplement  equation \eqref{ku3}=\eqref{ku4} with the initial condition
\be\label{in_cond}
u(-T)=0, 
\ee
for some $0<T \le +\infty$, and in the spirit of WT decompose a solution of \eqref{ku4}, \eqref{in_cond} to formal series in $\rho$:
\be\label{decomp}
	a(\tau)=a^{(0)}(\tau) +\rho a^{(1)}(\tau)+  \dots , \quad a^{(j)}(\tau) =a^{(j)}(\tau;\nu, L).
		\ee
 It can be easily seen that in the case $T=\infty$ the processes $a^{(j)}$ are stationary. 
At first, as in physical works (e.g. cf. \cite{Naz11}, Section~6.4) we retain the quadratic in $\rho$ part of this decomposition.
We denote
it $A_s(\tau)$, cal it a {\it quasisolution} and study its energy spectrum $N_s(\tau) = \EE(|A_s(\tau)|^2)$. In Sections~2-4 we show that
in order to obtain a nontrivial  asymptotic of  $N_s(\tau) $  the right scaling for the constant $\rho$ in eq.~\eqref{ku4} is
$\rho\sim\nu^{-1/2}$ and accordingly substitute in the equation $\rho = \nu^{-1/2}\eps^{1/2}$, where  $0<\eps\le1$
should be regarded as a small constant. 
Then $N_s$ may be written as
$
N_s(\tau) = N_s^0(\tau) +\eps N_s^1(\tau) +O( \eps^{2} ),
$
where $N_s^0, N_s^1\sim1$, 
uniformly in $\nu$ and $L$. Next in Theorem~\ref{t4} we prove that for $L\ge \nu^{-2-\epsilon}$ (cf.  \eqref{assumption}),  the function $s\mapsto N_s$ naturally 
extends to a function on $\R^d$, which is  $\eps^2$--close to a solution $m_s(\tau)$  of the damped/driven wave
 kinetic equation (WKE) 
\be\lbl{i WKE}
\dot m_s(\tau) =-2\ga_s m_s(\tau) +2b_s^2 +\eps K_s(m(\tau)), \quad s\in\R^d, 
\qquad m(-T) =0, 
\ee
for any $\tau\geq -T$, where $K_s$ is the wave kinetic integral (see \eqref{kin int}). 
In the last Section 5 we return to the complete decomposition \eqref{decomp} of solutions 
$a_s(\tau)$.  Accordingly, we decompose the energy spectrum of $a_s$ as 
\be\label{N_s}
n_s(\tau) = n^0_s(\tau)  +\rho n^1_s(\tau)  +\dots.
\ee
and  analyse this decomposition, scaling as before $\rho = \nu^{-1/2}\eps^{1/2}$.  

 Since characteristic time in our system is $\tau\sim 1$ and the slow time $\tau$ is defined as $\tau=\nu^{-1}t$, under the scaling $\rho\sim\nu^{-1/2}$  we have $t\sim \nu^{-1}\sim (size \ of \ nonlinearity)^{-2}$. This time scale coincides with that usually considered by physicists.

The kinetic limit, presented in  Section~\ref{s_kinetic},
 applies to quasisolutions of eq.~\eqref{ku3} and we are not sure that the result remains true for exact solutions of the equation. Still we believe that the 
 result and the method of its proof is valid for exact solutions of some other models of WT, and we will
 clarify this in the nearest future. 
 In this connection let us emphasize that in physical works, devoted to the subject, the WKE is always deduced for quasisolutions (that is, for energy spectrum corresponding to quadratic truncations of the formal series for solutions in amplitude) but not for the energy spectrum of solutions.

Results of Sections 2-4 are proved in \cite{DK} and those of Section 5 -- in \cite{DK2}. More discussion of the obtained results
maybe found in \cite{DK}.
All constants in this work do not depend on $\nu, L, \rho, \eps$  and $\tau, T$, unless the dependence 
is explicitly indicated. 
\smallskip

\medskip

\noindent 
{\bf Acknowledgments.} AD was supported by RFBR according to the research project 18-31-20031, and SK --  by the grant  18-11-00032 of Russian Science Foundation.

	\section{ Solutions  as formal  series in $\rho$. }\label{s_series}
	
	As in the introduction, let us  decompose a solution $a_s(\tau)$  of \eqref{ku4}, \eqref{in_cond} as formal series 
	\eqref{decomp}. Then 
$$
\dot a^{(0)} +\gamma_s a^{(0)}= b_s \dot\beta_s, \qquad  a^{(0)}(-T) =0,
$$
so $a^{(0)}$ is the Gaussian process 
	$
	a^{(0)}_s(\tau) = b_s \int_{- T}^\tau e^{-\gamma_s(\tau-l)}d\beta_s(l),
	$
	while $a^{(1)}$ satisfies
	$$
	\dot a^{(1)}_s (\tau)+ \gamma_s a^{(1)}_s (\tau)
	=  i \cY_s(a^{(0)}(\tau);\nu^{-1}\tau),\qquad a^{(1)}(-T) =0\,,
	$$
	so
	\be\label{a_s_1}
	a^{(1)}_s(\tau) = i \int_{- T}^\tau e^{-\ga_s(\tau-l)} \cY_s(a^{(0)}(l); \nu^{-1}l)\,dl
	\ee
 is a Wiener chaos of third order.  
Similar, for $n\ge2$, 
$$
\dot a^{(n)}_s (\tau)+ \gamma_s a^{(n)}_s (\tau)
	=  i  \sum_{n_1+n_2+n_3=n-1} \cY_s( a^{(n_1)}(\tau), a^{(n_2)}(\tau),  {a}^{(n_3)}(\tau) ;\nu^{-1}\tau \big),
$$
so
	\begin{equation} \label{so}
	\begin{split}
	a^{(n)}_s(\tau) 
	&= i \int_{- T}^\tau \\
	&\sum_{n_1+n_2+n_3=n-1} \,
	 e^{-\ga_s(\tau-l)}  \cY_s( a^{(n_1)}(l), a^{(n_2)}(l),  {a}^{(n_3)}(l) ;\nu^{-1}l \big)
	 \,dl
	\end{split}
	\end{equation}	
is a Wiener chaos of order $2n+1$.  We can iterate in the Duhamel integral in the r.h.s. of \eqref{so}
and eventually 
 express  $ a^{(n)}(l), l\ge -T$ via the processes $ a^{(0)}(l')$,  $l'\le l$.
 \medskip
 
To analyse the limiting behaviour of  correlations of solutions $a_s(\tau)$ and that 
of the energy spectrum $n_s(\tau)$, written as formal series \eqref{decomp} and \eqref{N_s}, we 
 should analyse the  limiting correlations of the processes $a^{(n)}_s(\tau)$. To give an idea what we should expect there, 
let us assume for a moment that $T=\infty$ and consider correlations of  $a^{(n)}_s(\tau_1)$ and $a^{(n')}_{s'}(\tau_2)$ with 
$n, n'\le1$. Then obviously, 
	\be\label{N0}
	\EE a_s^{(0)}(\tau_1)  a_{s'}^{(0)}(\tau_2) \equiv 0, \quad
	\EE a_s^{(0)}(\tau_1) \bar a_{s'}^{(0)}(\tau_2)
	= \delta^s_{s'} \,e^{-\gamma_s|\tau_1-\tau_2|} \frac{b_s^2}{\gamma_s}\,;
	\ee
it also can be shown that $\EE a_s^{(0)}(\tau_1)  a_{s'}^{(1)}(\tau_2) \equiv \EE a_s^{(0)}(\tau_1) \bar  a_{s'}^{(1)}(\tau_2) 
 \equiv0$. Denote 	
$\ 
B(s) = b_s^2/\ga_s,$ $ s\in \R^d.
$
Then, in view of \eqref{a_s_1}, \eqref{N0} and the Wick theorem
 the correlations of the processes $a_s^{(1)}(\tau)$  are given by 
\begin{equation*}
	\begin{split}
	\EE a_s^{(1)}(\tau)  a_{s'}^{(1)}(\tau) =0,& \quad
	\EE a_s^{(1)}(\tau) \bar a_{s'}^{(1)}(\tau) =\delta^s_{s'} J_s,  \\
	&J_s
	=\frac{2\nu^2}{\ga_s}
	L^{-2d}\sum_{1,2}\dess   \,\frac{\ga_{123s}\, B(s_1, s_2, s_3)}{(\oms)^2  +(\nu  \ga_{123s})^2}\,, 
		\end{split}
	\end{equation*}
	where 
	$$
	 \ga_{123s} = \ga_{1}+\ga_{2}+\ga_{3}+\ga_s, \qquad B(s_1, s_2, s_3) = B(s_1) B(s_2) B(s_3)
	 $$
(see \cite{DK} for the calculation). 
	  The sums $J_s$ may be well approximated by the integrals 
	$$
I_s =\frac{2\nu^2}{\ga_s}\!
		\int_{\R^d\times\R^d}  ds_1\,ds_2\, 
		\frac{  \ga_{123s}\,  B(s_1, s_2, s_3) }{ 4((s_1-s)\cdot (s_2-s))^2 +( \nu  \ga_{123s} )^2}, \quad
		s_3=s_1+s_2-s.
$$
Namely, 
\be\lbl{int approx}
|J_s -I_s| \le C^\#_sL^{-2}\nu^{-2} \quad \forall\, s. 
\ee
 Here and below by $ C^\#_s$ we denote various continuous functions of $s$ which decay, as $|s|\to\infty$, 
 faster than any  negative degree of $|s|$. Due to  \eqref{assumption} the r.h.s.  is small: it is bounded by $ C^\#_s\nu^{2+2\epsilon}$. 
 
 The asymptotical behaviour of integrals $I_s$ is known (see \cite{K, DK}): 
 
 \begin{theorem}\label{t1}
 We have $I_s =\nu I_s^0 +O( C_s^\# \nu^2)$,\footnote{If $d=2$, the term $O( C_s^\# \nu^2)$ should be replaced by $O( C_s^\#(\aleph) \, \nu^{2-\aleph})$, for any $\aleph>0$. }
    where 
 $$
  I_s^0=
		\frac{\pi }{\ga_s} \int_{\Sigma_s} 
		\frac{ B(s_1, s_2, s_1+s_2-s) }{{\sqrt{|s_1-s|^2+|s_2-s|^2}} \ }
		\, ds_1ds_2\!\!\mid_{\Sigma_s}.
 $$
 Here 
 $\Sigma_s$ is the quadric  $ \{(s_1, s_2): (s_1-s) \cdot (s_2-s)=0\}$ 
 and $ds_1ds_2\!\!\mid_{\Sigma_s}$ is the volume element  on it, corresponding to the Euclidean 
 structure on $\R^{2d}$. 
 \end{theorem}
 
   Asymptotic similar to that in Theorem~\ref{t1} can be obtained for arbitrary $-\infty\leq T<0$.
 Substituting $s_1=s+x, s_2= s+y$ and denoting $z=(x,y)$ we re-write $I_s^0$ as 
 $$
 \frac{\pi }{\ga_s} \int_{\Sigma} 
		\frac{ B(s+x, s+y, s+x+y) }{| z| } 
		\, dz\!\!\mid_{\Sigma}, \quad \Sigma = \{z=(x,y): x\cdot y=0\}.
 $$
 Denote $F(z) := x\cdot y = -\tfrac12 \oms\mid_{s_1=s+x, s_2=s+y, s_3=s+x+y}$. 
 Then $|\nabla F(z)| = |z|$, so the integral above  is exactly the integral $  \int B\delta(F)$ of the function   $B$  against  the delta-function of $F$,
 see \cite{ZLF}, p.~67. Since
$\ 
F=  -\tfrac12 \omega^{s+x\  s+y}_{s+x+y\  s} =  -\tfrac12\, \oms \des
$,
then neglecting the minus-sign we write $I^0_s$ as 
$$
  \frac{\pi }{\ga_s} \int B\,\delta(F)dxdy = \frac{2\pi }{\ga_s} \int  B\,\delta( \omega^{s+x\ s+y}_{s+x+y\ s} )dxdy =
 \frac{2\pi }{\ga_s} \int  B\,\delta( \oms) \, \des\, ds_1 ds_2 ds_3.
$$

Regarding $|z|^{-1} \, dz\!\!\mid_{\Sigma}=\delta(F)$ as a measure in the space $\R^{2d}$, supported by the quadric $\Sigma$, we show in \cite{DK} that it 
  may be disintegrated  as 
 $
 |x|^{-1} dx\,d_{x^\perp}y,
 $
 where $d_{x^\perp}$ is the Lebesgue measure on the space $x^\perp = \{y: y\cdot x=0\}$. That is, for a function $f(z)$ we have 
 \be\label{disint}
 \int_\Sigma f(z) |z|^{-1}  dz\!\!\mid_{\Sigma} = \int_{\R^d} |x|^{-1} \Big( \int_{x^\perp} f(x,y) d_{x^\perp}y\Big)  dx. 
 \ee
 This representation is instrumental to work with integrals of the form  $\int f\delta(F)$ and  is used below.

 \section{Quasisolutions  }\label{s_quasisol}
 As it was announced in Introduction, we  start with investigating   the quadratic in $\rho$ part of \eqref{decomp} which we call 
 a  quasisolution $A(\tau)= (A_s(\tau;\nu, L), s\in \Z^d_L)$: 
 $$
 A_s(\tau) = a^{(0)}_s(\tau) +\rho a^{(1)}_s(\tau)+ \rho^2a^{(2)}_s(\tau).
 $$
 Consider the energy spectrum of   $A$, 
 $
 N_s(\tau) = \EE |A_s(\tau)|^2,
 $
 and decompose it in series in $\rho$: 
  \be\label{ns}
 N_s(\tau;\nu, L) =  n_s^0(\tau)  +\rho  n_s^1(\tau)  +\rho^2  n_s^2(\tau) + \rho^3  n_s^3(\tau) +\rho^4  n_s^4(\tau) , 
 \ee
 where 
 $ n_s^j(\tau) = n_s^j(\tau; \nu,L)$. 
 Here 
 $ n_s^0 = \EE|a^{(0)}_s(\tau)|^2 \sim C_s^\#
 $,\footnote{We write $m_s \sim C_s^\#$ if $m_s\le C_s^\#$ for all $s$ and $\|m\| \ge C^{-1}$ (see \eqref{norm}), uniformly in $\nu$.  } 
 and it is easy to see that $ n_s^1\equiv0$. 
 Next, $ n_s^2 = \EE |a^{(1)}_s|^2 + 2\Re   \EE a^{(0)}_s \bar a^{(2)}_s$. The first term $\EE |a^{(1)}_s|^2$  is of order $\nu$
 and  is given by Theorem~\ref{t1} if $T=\infty$; 
 the second one is similar. So
 \be\label{n_s_od}
  n_s^0   \sim C_s^\#,\; n_s^1\equiv 0,\; 
 n_s^2 \sim C^\#_s \nu. 
 \ee
  It turns out that \footnote{ If $d=2$, the estimate for $n_s^3$ should be replaced by
  $
  | n_s^3| \le  C^\#_s \nu^2\ln\nu^{-1}.
  $}
 \be\label{N34}
  |n_s^3|, | n_s^4| \le C^\#_s\, \nu^2,
 \ee
 if $L\ge \nu^{-2-\epsilon}$ (see \eqref{assumption}). 
 
 For any $\tau\ge -T$, any $\nu, L$ and any $k=0, \dots, 4$ the function $s\mapsto n_s^k(\tau)$ naturally extends to a Schwartz function on $\R^d$. The limit 
 $$
 n_s^k(\tau;\nu, \infty) = \lim_{L\to\infty} n_s^k(\tau;\nu, L)
 $$
 exists, is a Schwartz 
 function of $s\in\R^d$ and satisfies \eqref{n_s_od}, \eqref{N34}. Accordingly, the limiting energy spectrum of a quasisolution,
 $N_s(\tau; \nu, \infty)= \lim_{L\to\infty} N_s(\tau;\nu, L)$, also exists and  is a  Schwartz function of $s\in\R^d$.
 \smallskip
 
 Relations \eqref{n_s_od} and \eqref{N34} suggest that the right scaling for $\rho$ is 
 $
 \rho\sim \nu^{-1/2},
 $
 and  we choose  $\rho$ to be of the form
  \be\label{rho}
 \rho= \nu^{-1/2} \eps^{1/2}, \quad \eps\in (0, 1]. 
 \ee
 With this choice of $\rho$  the process $N_s(\tau)$ is an $\eps$-small perturbation of the liner 
 process $n_s^0$ and does not converge to $n_s^0$ under the limit \eqref{assumption}: decomposition  \eqref{ns} takes the form
 $N_s=n_s^0 + \eps N_s^1 + O(\eps^2),$
 where	$N_s^1 = \nu^{-1}n_s^{2}
 \volna C^\#_s$.

  \section{The wave kinetic equation  }\label{s_kinetic}
 
For a real function $\R^d\ni s \mapsto y_s$ consider the corresponding  cubic wave kinetic integral: 
 \be\lbl{kin int}
  K_s(y_\cdot) = 2\pi
  \int_{\Sigma_s} \frac{ds_1\, ds_2\!\mid_{\Sigma_s}\, y_{1}y_{2}y_{3}y_{s}}{ \sqrt{|s_1-s|^2+|s_2-s|^2}}\left(
\frac1{ y_{s}} +\frac1{y_{3}} -\frac1{ y_{1}}-\frac1{ y_{2}}
\right),
 \ee
 where for $j=1,2,3$ we denote 
  $y_j=y_{s_j}$ and where   $s_3 = s_1+s_2-s$. 
 Using the notation, evoked after  Theorem~\ref{t1}, the integral above may be written as
 $$
   4\pi
  \int
  y_{1}y_{2}y_{3}y_{s} 
  \left(
\frac1{ y_{s}} +\frac1{ y_{3}} -\frac1{ y_{1}}-\frac1{ y_{2}}
\right) \,\delta( \oms) \, \des\,ds_1 ds_2 ds_3. 
 $$
 This  integral exactly coincides with the kinetic integral, used  by physicists  to describe 
   WT for the 4--waves 
 interaction, see \cite{ZLF}, p.71 and \cite{Naz11}, p.~91.

  Consider the function spaces \ 
 $
 C_r(\R^{d}) =\{x\in C (\R^{d}) : |x|_r = \sup_s (1+ |s|)^r |x(s)| <\infty \}.
 $
 The representation \eqref{disint} for the measure  $|z|^{-1}  dz\!\!\mid_{\Sigma}$ implies that the wave kinetic integral
 $K$ defines a continuous operator 
 $$
 K: C_r(\R^{d}) \to C_{r+1}(\R^{d}), \qquad y_s \mapsto K_s(y_\cdot),
 $$
 provided that   $ r>d$. 
Now let us consider the wave kinetic equation:
\be\label{w_k_e}
\dot  m_s (\tau) = -2\ga_s  m_s (\tau) +2b_s^2 + \eps \, K_s(m_\cdot(\tau)),\quad s\in \R^d.
\ee
There exists $\eps_*>0$ 
 such that if  $0\le\eps\le \eps_*$, then \eqref{w_k_e} 
 has a unique solution $m^*(\tau)$, vanishing at $\tau=-T$ and defining 
  a bounded continuous curve  $m^* :[-T,\infty) \mapsto C_r(\R^d)$, in any space $C_r(\R^d)$. 
It may  be written as 
$$
m_s^*(\tau ) =  m_s^{*0}(\tau ) + \eps m_s^{*1}(\tau;\eps ), \qquad m^{*0}(-T)= m^{*1} (-T)=0, 
$$ 
where $m^{*0}, m^{*1} \sim1$, 
  $m_s^{*0}(\tau)$ equals $n_s^0(\tau)$ and 
  satisfies the linear equation 
 \be\label{linear}
 \dot m_s^{*0}(\tau) = -2\ga_s m_s^{*0}(\tau) + 2b_s^2. 
 \ee
 
  Everywhere below $\eps$ is  a fixed small constant, independent from $\nu$ and $L$, satisfying  $\eps \in(0,  \eps_*].$
 The parameter $\eps$  should be interpreted as the squared amplitude of a quasisolution $A(\tau) $, written in a  right scaling. 
  The following theorem is the main result of \cite{DK}:

\begin{theorem}\label{t4}
Let in \eqref{ku3} $L\ge \nu^{-2-\epsilon}$ and $\rho= \nu^{-1/2} \eps^{1/2}$. Then the energy spectrum $N_s=N_s(\tau;\nu, L)$ of a
 quasisolution $A_s$ is $\eps^2$-close to the solution $m^*$ of \eqref{w_k_e} in the sense that for any $r$
$$
|m^* (\tau) - N (\tau)|_r \le C_r \eps^2\qquad \forall\, \tau\ge-T, 
$$
with some $C_r>0$, provided that 
$0<\nu\le\nu_{\eps,r}$ for a suitable $ \nu_{\eps,r}>0$.  The limiting energy spectrum 
$N_s(\tau;\nu, \infty)$ also satisfies the estimate above for any $r$ and for 
$0<\nu\le\nu_{\eps,r}$.
\end{theorem}

Eq. \eqref{linear} has the unique steady state $m^0$, $m^0_s= b_s^2/\ga_s$, which  is asymptotically stable. By
the  implicit function theorem,  for $\eps$ sufficiently small
eq.~\eqref{w_k_e} has a unique steady state $m^\eps$ close to $m^0$, which is asymptotically stable. Decreasing $\eps_*$ if needed we 
may assume 
that the unique $m^\eps$ exists for $\eps\le\eps_*$. 
Jointly with Theorem~\ref{t4} this result describes the asymptotic in time behaviour of the 
energy spectrum $N_s(\tau)$:
for any  $r>d$,
\be\label{time_ass}
|m^\eps - N(\tau)|_r \le C_r \big(|m^\eps|_r e^{-\tau-T} +\eps^2 \big), \qquad \forall \tau\ge-T.
\ee

Due to Theorem \ref{t1}  together with \eqref{int approx} and some modifications of these results, the iterated limit
$ \ 
\lim_{\nu\to0} \lim_{L\to\infty} \nu^{-1} n_s^{2}(\tau; \nu, L) 
$
exists and is non-zero. It is hard to doubt that a similar iterated limit also exists for $ \nu^{-2} n_s^{4}$ 
(however, we have  not proved this yet). Then, in view of
estimates 
\eqref{N34}, under the scaling $\rho = \nu^{-1/2} \eps^{1/2}$ 
 exists the iterated  limit
$
N_s(\tau; 0,\infty)=   \lim_{\nu\to0} \lim_{L\to\infty} N_s(\tau; \nu,L).
$
If so, then the it 
 also satisfies the assertion of Theorem~\ref{t4} and the time-asymptotic \eqref{time_ass}.
  \medskip

In the fast time $t$  eq. \eqref{ku3} with $\rho$ as above and  $ \lambda = (\nu\,\eps)^{1/2} $
reeds 
$$
	u_t +   i  \Delta u -   i \lambda \,( |u|^2 -\|u\|^2) u 
	 =- \nu(-\Delta +1)^{r_*} u +  \sqrt\nu\, \dot \eta^\omega(\tau,x),  
$$
where $ \| u(t)\|\sim 1$ as $\nu\to0$, $L\to\infty$ 
by \eqref{balance}.  We have seen that

1) when $\nu\to0, L\to\infty$, 
the coefficient $\lambda$ in front of the nonlinearity should scale as $\sqrt\nu$ for the kinetic limit to exist.

2) The time, needed to arrive at the limiting
kinetic regime is $t\sim \lambda^{-2}$. The corresponding 
 characteristic time scale      $(\mbox{{\it size of the nonlinearity}})^{-2}$  coincides with the 
time scale usually considered by physicists, see \cite{Naz11, NR,  ZLF}.

  \section{Energy spectra of solutions $a_s(\tau)$, written  as formal series in $\rho$.  }\label{s_form_ser}

Let us come back to the decomposition \eqref{decomp}. 
As it was mentioned in Section \ref{s_series}, iterating the integral \eqref{so}
we may express each $a^{(n)}_s(l_0) $ via the processes $ a^{(0)}(l')$,  $l'\le l_0$.
Then $a^{(n)}_s$ represents as the sum
 \be\label{a_sum}
 a^{(n)}_s(l_0)  = \sum_{\cT\in\Ga(n)} I_s(l_0;n, \cT),
 \ee
 where the meaning of the summation index $\cT$ is explained below and  $ I_s(\cT) := I_s(l_0;n, \cT)$ is an iterated integral of the form 
\be\lbl{II1}
  I_s(\cT)
 = \int\dots\int L^{-nd}\sum_{s_1,\ldots,s_{3n}} (\dots) \,dl_1\dots dl_n.
\ee
 The integrating zone in \eqref{II1}  is a	 convex polyhedron in $[-T, l_0]^n$.
 The summation is taken over vectors $s_1\ldots,s_{3n}\in\Z^d_L$
 which are subject to the linear relations, 
 following from the factor $\dess$ in the definition of $\cY_s$. 
 The summand $ (\dots)$  in \eqref{II1}
 is a product of functions 
 $e^{-\gamma_{s'}(l_k-l_j)}$, $\exp({\pm i\nu^{-1} \om^{s'_1 s'_2}_{s'_3 s'_4}})$  and the processes $[a^{(0)}_{s^{''}}(l_r)]^*$, where $a^*$ is either $a$ or $\bar a$, 
 with various indices $k,j,r$ and various $s',s'_i,s''$, taken  from the set $\{s_1,\ldots, s_{3n}\}$. It is of degree $2n+1$ with respect to the process $a^{(0)}$. 
 Each integral $  I_s(l_0;n, \cT)$  corresponds to an oriented  rooted tree $\cT$ from a class 
 $ \Gamma(n)$ of trees with the root  at  $a^{(n)}_s$, with random 
 variables $[a^{(0)}_{s^{'}}(l_j)]^*$ at its leaves, and with
   vertices labelled by symbols  $[a^{(n')}_{s^{'}}(l_r)]^*$ with $1\le n'<n$,     see fig.~\ref{f:tree}.  To
 each vertex enters one edge of the tree and three edges outgo from it. For a vertex, labelled by some  
 $a^{(\bar n)}_{s'}(l')$, $\bar n\ge1$, 
 the three edges outgo to the vertices, corresponding to a  
  choice of the three terms $a^{(n_1)}, a^{(n_2)} , a^{(n_3)}$ in the 
 decomposition \eqref{so} of $a_s^{(n)}(\tau) :=
 a^{(\bar n)}_{s'}(l')$. 
 
 \begin{figure}[t]
 	\begin{center}\parbox{3.5cm} {
 		\begin{tikzpicture}[]
 		\node at (-1.4,0) (c0) {$a_{s}^{(2)}(\tau)$};
 		
 		\node at (-2.7, -1.2) (c1) {$a_{s_1}^{(0)}(l_1)$};
 		\node at (-1.4, -1.2) (c2) {$a_{s_2}^{(0)}(l_1)$};
 		\node at (-0.1, -1.2) (c3) {$\bar a_{s_3}^{(1)}(l_1)$};
 		
 		\node at (-2.7, -2.5) (c4) {$a_{s_4}^{(0)}(l_2)$};
 		\node at (-1.4, -2.5) (c5) {$\bar a_{s_5}^{(0)}(l_2)$};
 		\node at (-0.1, -2.5) (c6) {$\bar a_{s_6}^{(0)}(l_2)$};
 		
 		\draw [line width=0.25mm] (c0.south)--(c1.north);
 		\draw [line width=0.25mm] (c0.south)--(c2.north);
 		\draw [line width=0.25mm] (c0.south)--(c3.north);
 		\draw [line width=0.25mm](c3.south)--(c4.north);
 		\draw [line width=0.25mm](c3.south)--(c5.north);
 		\draw [line width=0.25mm](c3.south)--(c6.north);		
 		\end{tikzpicture}
 	}
 \end{center}
 \caption{A tree $\cT$ from the set $\Ga(2)$.}
 \lbl{f:tree}
\end{figure}

Accordingly we write the energy spectrum of a solution $a$ 
 as formal series \eqref{N_s}. There 
$n_s^0\sim1$, $n_s^1=0$ and $n_s^2$ are the same as in \eqref{ns}, but $n_s^3$ and $n_s^4$ are different;
this small ambiguity should not cause a  problem. The new coefficients  $n_s^3$ and $n_s^4$ still meet 
\eqref{N34} (see below). 
 Let us consider any $n_s^k(\tau)$, $k\ge0$. It equals 
$
\ n_s^k(\tau) = \EE \sum_{k_1+k_2=k}  a_s^{(k_1)}(\tau) \bar a_s^{(k_2)}(\tau),
$
where each  $ a_s^{(k)}(\tau)$ is given by the finite sum \eqref{a_sum},  parametrised by the trees
$\cT\in \Gamma(k)$. So 
$$
 \EE  a_s^{(k_1)}(\tau) \bar a_s^{(k_2)}(\tau) =\sum_{\cT_1\in \Gamma(k_1), \cT_2\in \Gamma(k_2)} \EE I_s(\tau;k_1, \cT_1) \overline{I_s(\tau;k_2, \cT_2)}.
$$
Here
\be\lbl{II}
\EE I_s(\cT_1) \overline{I_s(\cT_2)} = \int\ldots\int L^{-kd}\sum_{s_1,\ldots, s_{3k}}
\EE(\ldots)\,dl_1\ldots dl_{k},
\ee
where in the brackets under   the expectation sign stands a product of the terms in the 
 brackets from the representations  \eqref{II1} for  integrals $I_s(\cT_1)$ and $\overline{I_s(\cT_2)}$.
Since $ a_s^{(0)}(l), s\in \Z^d_L$, are  Gaussian random variables whose correlations are given by \eqref{N0} (the second correlation should be modified if $T< \infty$), then by the Wick formula each expectation $\EE  I_s(\cT_1) \overline{I_s(\cT_2)}$ is a (finite) sum over different Wick-pairings of non-conjugated variables $a_{s_j}^{(0)}(l_m)$ with conjugated variables $\bar a_{s_r}^{(0)}(l_q)$.
Since these  variables  correspond to  leaves  of the tree $\cT_1$ or $\ov\cT_2$, then
the  summands in \eqref{II}  can be parametrised by  Feynman diagrams,  obtained by 
paring the trees $\cT_1$ and $\ov\cT_2$  via the Wick-coupled leaves. 
\footnote{The coupled leaves may both belong to 
	$ \cT_1$ or to $ \bar \cT_2$. 
}
As for $s'\neq s''$ the Gaussian variables $a_{s'}^{(0)}$ and $\bar a_{s''}^{(0)}$ are uncorrelated, then 
 the summation $\sum_{s_1 ,\ldots, s_{3k}}$ in \eqref{II} is taken only over those vectors $(s_1,\ldots,s_{3k})$ for which  
 all Wick-paired variables $a_{s'}^{(0)}$ and $\bar a_{s''}^{(0)}$ have equal indices $s'=s''$. 
Thus, in every Feynman diagram for \eqref{II}  each leaf $a_{s}^{(0)}(l_m)$ of
$\cT_1$ is paired either with a leaf  $\bar a_{s}^{(0)}(l_q)$ of  $\cT_1$, or with a leaf 
$\bar a_{s}^{(0)}(l_q)$ of   $\overline{\cT_2}$, etc. 
We have seen that 
\begin{equation}\label{F}
n^k_s(\tau) =\sum_{\cF\in\gF(k)} \I_s(\tau;k,\cF), 
\end{equation}
where the sum is taken over the set $\gF(k)$ of Feynman diagrams, associated to 
all possible pairings of the trees $\cT_1\in\Ga(k_1)$ and $\ov\cT_2\in\ov\Ga(k_2)$, $k_1+k_2=k$, via their leaves. 

Resolving all the restrictions, imposed on  the indices $s_j$ in \eqref{II}   by an appropriate affine transformation, 
we find that among the $3k$ indices $s_j$ exactly $k$  are linearly  independent. 
Denoting by $z_1,\ldots,z_k\in\Z^d_L$ the independent variables obtained from the indices $s_j$ by this transformation, we write 
the sum in \eqref{II} as  
$L^{-kd}\sum_{z_1,\ldots,z_k\in\Z^d_L}$. Approximating the latter by an integral $\int_{\R^{kd}}\dots\,dz$ where $z=(z_1,\ldots,z_k),$
we get  for the integrals  $\I_s(\tau;k,\cF)$ 
a rather simple explicit formula. Namely, for any $s\in\Z^d_L$, 
\be\label{I_F}
	\begin{split}
	&\I_s(\tau;k,\cF) = J_s(\tau;k,\cF) + O\big(L^{-2}\nu^{-2}C_s^\#(k)\big),\\
&J_s(\tau;k,\cF) = \int_{\R^k}dl \,\int_{\R^{kd}}dz \,F_{\cF}(\tau,s,l,z) e^{i\nu^{-1}\sum_{i,j=1}^k \al^\cF_{ij}(l_i-l_j)z_i\cdot z_j} ,
		\end{split}
	\ee
where $l=(l_1,\ldots, l_k)$, the function $F_\cF$ is smooth in $(s,z)\in \R^{(k+1)d} $ and fast decays in $s,z$ and $l$,
while $\al^\cF=(\al^\cF_{ij})$ is a skew-symmetric  (constant) matrix without zero lines and rows. Its rank is at least two, and for some diagrams $\cF$ 
it equals two.  Moreover, each function $s\mapsto \I_s(\tau;k,\cF)$ naturally extends to a Schwartz function on $\R^d$, and after
this extension \eqref{I_F} holds for every $s\in \R^d$. Consequently, for any fixed $\nu>0$, 
$$
 \lim_{L\to\infty} \I_s(\tau;k,\cF) = J_s(\tau;k,\cF) \quad \forall \tau\ge-T,\ 
 \,s\in\R^d,\  k\ge1, \  \cF\in\gF(k).
$$
Then in view of \eqref{F} we have
$$
n_s^k(\tau; \nu, \infty) := \lim_{L\to\infty} n_s^k(\tau;\nu, L) = \sum_{\cF\in\gF(k)} J_s(\tau;k,\cF), 
$$
for all $k,s$ and $\tau$.

 Relations \eqref{n_s_od} and \eqref{N34} suggest to assume that 
 \be\label{assume}
  |n_s^k|\le C^\#_s(k) \nu^{k/2 } \qmb{for all}\qu \nu,
  \ee
  for every $k$, 
  if $L$ is sufficiently big in terms of $\nu^{-1}$ and $k$. 
  In this direction we have   the   two theorems below:

\begin{theorem}\label{t2}
 For each $k$ and each $\cF\in\gF(k)$, \footnote{If $k=3$ and $d=2$, then the r.h.s. below should be modified by the factor
 $ \ln \nu^{-1}$.}
\be\label{A_est}
  | J_s(\tau;k,\cF) |
 \le C^\#_s(k) \nu^{\min(\lceil k/2\rceil, d)} , \quad \forall\, \tau\ge-T, 
\ee
where  ${\lceil k/2  \rceil} $ is the smallest integer $\ge k/2$.
\end{theorem}

By \eqref{I_F}  if $L$ is so big that 
\be\label{L_big}
L^{-2} \nu^{-2} \le \nu^{\min(\lceil k/2\rceil, d)} ,
\ee
then $\I_s$ also satisfies \eqref{A_est}, and in view of \eqref{F}  $n_s^k(\tau)$  is bounded by the r.h.s. of \eqref{A_est}, multiplied by $|\gF(k)|$.  So 
\eqref{assume} holds true for $k\le4$ since $d\ge2$. 
But for $k$ large in terms of $d$ the upper estimate  \eqref{A_est}  is worse 
   then the desired bound  \eqref{assume}, and our next result shows that  estimate \eqref{A_est} is sharp in the sense
   that  in the exponent in the r.h.s.  of  \eqref{A_est}, ${\min(\lceil k/2\rceil, d)}$ cannot be replaced by $\lceil k/2\rceil$.

  Let $\gF_2(k)\subset\gF(k)$ be a set of Feynman diagrams $\cF$, 
   for which the matrix $\al^\cF$ from \eqref{I_F}  has exactly one nonzero row and column.  This set is not empty.

\begin{theorem}\label{t3}
 If $k > 2d$, then for any $\cF\in\gF_2(k)$ we have $ J_s(\tau;k,\cF)  \sim \nu^d C_s^\#(k) \gg \nu^{\lceil k/2\rceil }C_s^\#(k)$.
 But in the same time, 
 \be\label{cancel}
\Big|\sum_{\cF\in\gF_2(k)}  J_s(\tau;k,\cF) \Big|
 \le \nu^{k-1} C_s^\#(k) \ll \nu^{\lceil k/2\rceil }C_s^\#(k).
\ee
\end{theorem}

It is plausible that  the  cancellation, leading to  the validity of \eqref{cancel}, is a general fact, and we suggest the following problem:

 \begin{problem}\label{prb}
 Prove that $\Big|\sum_{\cF\in\gF(k)}  J_s(\tau;k,\cF) \Big| \leq C^\#_s(k)\nu^{k/2}$  for all $k$ and all $\nu$.   In particular, 
$
	|n_s^k(\tau; \nu, \infty)| \le C_s^\# (k) \nu^{ k/2}
$
and \eqref{assume} holds if $L$ is sufficiently big in terms of $\nu^{-1}$ and $k$. 
\end{problem}

If this conjecture is true, then  under the substitution \eqref{rho} the limiting  decomposition 
$\ 
n_s(\tau;\nu, \infty) = n^0_s(\tau;\nu, \infty) +\rho n^1_s(\tau;\nu, \infty) +\dots
$
becomes a formal series  in $\sqrt\eps$, uniformly in $\nu$. 
So for any $M\ge2$ its truncation of order $M$,
 $
 n_{s,M} (\tau;\nu, \infty)=   n^0_s(\tau;\nu, \infty)   +\dots + \rho^M n^M_s(\tau;\nu, \infty),
 $
 is $\eps^2$--close to $N_s(\tau;\nu, \infty)$ and also meets the assertion of Theorem~\ref{t4}. It is unclear for us if for a large $M$ the truncation
   $ n_{s,M}(\tau)$ satisfies equation 
  \eqref{w_k_e} with  an accuracy, better  than $\eps^2$.
  
  On the contrary, if the conjecture in Problem \ref{prb}    is wrong in the sense that 
  $$
  \sup_{\tau\ge -T} \| n^k(\tau;\nu, L)\| \ge C \nu^{k/2-\kappa}, \quad \kappa>0, 
  $$
  for some $k$\ \footnote{\  It must be $\ge5$ by \eqref{N34}.}
  and for all sufficiently small $\nu$ and large $L$, 
   then  \eqref{N_s} with $\rho=\nu^{-1/2} \eps^{1/2}$ 
    is not a formal series  in $\sqrt\eps$ uniformly in 
  $\nu$. 
   We do not rule out this possibility since  NLS equations appear in physics as models for small oscillations in various media, obtained by neglecting in the 
  exact equations terms of  high order in the amplitude. So it is not impossible that the kinetic limit holds for the energy spectra
  of quasisolutions, but not for  the exact  energy spectrum or for the energy spectrum of high order in $\rho$ truncations of the series 
  \eqref{decomp}.

\end{document}